\documentclass[12pt]{article}
\usepackage{epsfig}
\usepackage[T1]{fontenc}
\usepackage{graphics}

\usepackage[latin1]{inputenc}

\usepackage[english,french]{babel}

\usepackage{graphicx}

\usepackage{dcolumn}

\usepackage{moreverb}

\usepackage{amsmath,amssymb,amsfonts}

\begin{document}
\numberwithin{equation}{section}

\begin{flushleft}
Journal of Mathematical Physics
\newline
\newline
\newline
\title*{\textbf{Polynomial Poisson Algebras for
Classical Superintegrable Systems with a Third Order Integral of
Motion}}
\newline
\newline
Ian Marquette
\newline
D\'epartement de physique et Centre de recherche math\'ematiques,
Universit\'e de Montr\'eal,
\newline
C.P.6128, Succursale Centre-Ville, Montr\'eal, Qu\'ebec H3C 3J7,
Canada
\newline
ian.marquette@umontreal.ca
\newline
514-343-6111 (4072)
\newline
\newline
Pavel Winternitz
\newline
D\'epartement de math\'ematiques et de statistique et Centre de
recherche math\'ematiques,Universit\'e de Montr\'eal,
\newline
C.P.6128, Succursale Centre-Ville, Montr\'eal, Qu\'ebec H3C 3J7,
Canada
\newline
wintern@CRM.UMontreal.CA
\newline
514-343-7271
\newline
\newline
We present polynomial Poisson algebras for the 8 classical
potentials in two-dimensional Euclidian space that separate in
cartesian coordinates and allow a third order integral of motion.
Some of the classical superintegrable potentials do not coincide
with quantum ones, but are their singular limits. We present the
trajectories for all these classical potentials. We find that all
bounded trajectories are periodic.
\newline
\newpage
\section{Introduction}
The purpose of this article is to study the properties of a certain
class of superintegrable systems in a real two-dimensional Euclidian
space $E_{2}$. These are classical hamiltonian systems that allow
three functionally  independent integrals of motion. One of them is
the Hamiltonian H. The other two are second and third order
polynomials in the momenta, respectively. Moreover, in most of the
article we shall restrict to the case when the potential allows
separation of variables in Cartesian coordinates, so that we have
\newline
\begin{equation}
H = \frac{1}{2}(P_{1}^{2}+P_{2}^{2})+f(x)+g(y)
\end{equation}
\begin{equation}
A =\frac{1}{2}(P_{1}^{2}-P_{2}^{2})+f(x)-g(y)
\end{equation}
\newline
and $X_{3}$ will have the general form
\newline
\begin{equation}
X_{3} =
\sum_{i+j+k=3}a_{ijk}L_{3}^{i}P_{1}^{j}P_{2}^{k}+k_{1}(x,y)P_{1}+k_{2}(x,y)P_{2},
\end{equation}
where $a_{ijk}$ are constant and $L_{3}$ is the angular momentum
\begin{equation}
L_{3} = xP_{2}-yP_{1}.
\end{equation}
Superintegrable systems are Hamiltonian systems with more integrals
of motion than degrees of freedom. In dimension n=2 the maximal
possible number of functionally independent integrals is 2n-1=3 and
that is the case we are considering.
\newline
\newline
A large body of literature exists on superintegrable systems,
inspired by the superintegrability of the harmonic
oscillator$^{1,2}$ and Kepler-Coulomb potentials$^{3,4}$. A
systematic search for superintegrable systems in $E_{2}$ and $E_{3}$
was conducted in the 1960ties$^{5,6,7}$. Like much of the later work
on superintegrable systems it was restricted to the case of second
order integrals of motion$^{8,..,13}$. This case turned out to have
an intimate connection with the separation of variables in the
Hamilton-Jacobi and Schrödinger equations.
\newline
\newline
An interesting discovery was that in general the quadratic integrals
of motion do not generate finite dimensional Lie algebras, but more
complicated algebraic structures, namely quadratic
algebras$^{14-18}$.
\newline
Much less is known about integrable and superintegrable systems with
third and higher order integrals of motion. In 1935 Drach found 10
different potentials in a complex Euclidian plane allowing a third
order integral of motion in classical mechanics$^{19,20}$. In 1984
Hietarinta$^{21}$ showed that in the case of integrals that are
third, or higher order polynomials in momenta, quantum and
classically integrable potentials may not coincide.
\newline
A systematic search for superintegrable systems involving at least
one third order integral was started in 2002$^{22}$. It was shown
that in the case of one first order and one third order integral the
only classical superintegrable systems were known ones (like
$V=\alpha r^{2}$ or $V=\alpha r^{-1}$) for which the third order
integral was the product of a first and second order one. However,
in the quantum case a new superintegrable potential of this type
exists, namely $V_{1}(x,y)=(\hbar \omega)^{2}k^{2}sn^{2}(\omega
x,k)$ where $sn(\omega x,k)$ is a Jacobi elliptic function. The
classical limit ($\hbar \rightarrow 0$) is free motion!$^{22}$
\newline
All classical and all quantum potentials allowing a second order
integral of the form (1.2) and a third order integral (of any form )
were found in ref.$^{23}$. The classical ones are all expressed in
terms of elementary functions. The quantum ones are quite different
and some of them involve elliptic functions and Painlev\'e
transcendents$^{24}$. We leave the quantum case for a future study
and concentrate here on the classical superintegrable Hamiltonian
systems of the form (1.1),...,(1.3)
\newline
\newline
In Section 2 we consider a general superintegrable Hamiltonian
system in a two-dimensional space (not necessarily Euclidian) with a
scalar potential. It allows one quadratic and one cubic integral of
motion. We construct the most general cubic Poisson algebra
generated by these integrals. We express the Casimir operator K of
this algebra in terms of the Hamiltonian H. The general formalism of
Section 2 is applied in Section 3 to the 8 existing superintegrable
systems in $E_{2}$ allowing separation of variables in Cartesian
coordinates and having an additional third order integral of motion.
Solutions of the equation of motion and trajectories in these 8
systems are discussed in Section 4. Section 5 is devoted to
conclusions.
\newline
\section{Cubic Poisson algebras}
We start out with a more general Hamiltonian system than
(1.1)...(1.3). We will not assume that we have a natural Hamiltonian
in $E_{2}$, but that we have a superintegrable system with a
quadratic Hamiltonian and one second order and one third order
integral of motion. We put
\begin{equation*}
H = a(q_{1},q_{2})P_{1}^{2} + 2b(q_{1},q_{2})P_{1}P_{2} +
c(q_{1},q_{2})P_{2}^{2} + V(q_{1},q_{2})
\end{equation*}
\begin{equation*}
A = A(q_{1},q_{2},P_{1},P_{2})= d(q_{1},q_{2})P_{1}^{2} +
2e(q_{1},q_{2})P_{1}P_{2} +
\end{equation*}
\begin{equation}
f(q_{1},q_{2})P_{2}^{2} + g(q_{1},q_{2})P_{1} + h(q_{1},q_{2})P_{2}
+ Q(q_{1},q_{2})
\end{equation}
\begin{equation*}
B = B(q_{1},q_{2},P_{1},P_{2}) = u(q_{1},q_{2})P_{1}^{3} +
3v(q_{1},q_{2})P_{1}^{2}P_{2} + 3w(q_{1},q_{2})P_{1}P_{2}^{2}
\end{equation*}
\begin{equation*}
+x(q_{1},q_{2})P_{2}^{3}+j(q_{1},q_{2})P_{1}^{2}+
2k(q_{1},q_{2})P_{1}P_{2} + l(q_{1},q_{2})P_{2}^{2} +
m(q_{1},q_{2})P_{1} + n(q_{1},q_{2})P_{2} + s(q_{1},q_{2})
\end{equation*}
\newline
where $P_{i}$ and $q_{i}$ are canonical momenta and coordinates and
a,..,s are functions to be determined. Since A and B are integrals
of motion we have.
\newline
\begin{equation}
\{H,A\}=\{H,B\}=0
\end{equation}
\newline
where $\{,\}$ is the Poisson bracket
\newline
\newline
Since B is cubic in the momenta, we cannot expect $\{H,A$ and B$\}$
to generate a quadratic algebra, like those obtained
in$^{14,...,18}$. We will however try to close the algebra at the
lowest order possible, namely 3. We put
\newline
\[ \{A,B\}=C \]
\begin{equation}
\{A,C\}= \alpha A^{2} + 2\beta AB + \gamma A + \delta B + \epsilon
\end{equation}
\[ \{B,C\}= \mu A^{3} + \nu A^{2} + \rho B^{2} + 2\sigma AB + \xi A + \eta B + \zeta \]
\newline
Since we have $\{C,\{A,B\}\}$=0 the Jacobi identity reduces to
\newline
\begin{equation}
\{A,\{B,C\}\}=\{B,\{A,C\}\}
\end{equation}
\newline
The Jacobi identity implies
\newline
$\rho=-\beta , \sigma=-\alpha , \eta=- \gamma $ and we obtain the
cubic algebra.
\newline
\[ \{A,B\}=C \]
\begin{equation}
\{A,C\}= \alpha A^{2} + 2\beta AB + \gamma A + \delta B + \epsilon
\end{equation}
\[ \{B,C\}= \mu A^{3} + \nu A^{2} - \beta B^{2} - 2\alpha AB + \xi A - \gamma B + \zeta  .     \]
\newline
The coefficients $ \alpha$ , $\beta$ and $\mu$ are constants, but
the other ones can be polynomials in the Hamiltonian H. The degrees
of these polynomials are dictated by the fact that H and A are
second order polynomials in the momenta and B is a third order one.
Hence C can be a fourth order polynomial. We have
\newline
\begin{equation}
\alpha=\alpha_{0}, \beta=\beta_{0}, \mu=\mu_{0}
\end{equation}
\[\gamma=\gamma_{0}+\gamma_{1}H,\delta=\delta_{0}+\delta_{1}H,\epsilon=\epsilon_{0}+\epsilon_{1}H+\epsilon_{2}H^{2}\]
\[\nu=\nu_{0}+\nu_{1}H,\xi=\xi_{0}+\xi_{1}H+\xi_{2}H^{2}\]
\[\zeta=\zeta_{0}+\zeta_{1}H+\zeta_{2}H^{2}+\zeta_{3}H^{3},\]
\newline
where $\alpha_{0}$,...,$\zeta_{3}$ are constants. A Casimir operator
K of a polynomial algebra is defined as an operator Poisson
commuting with all elements of the algebra. For the algebra (2.5)
this means
\newline
\begin{equation}
\{K,A\}=\{K,B\}=\{K,C\}=0
\end{equation}
\newline
and this implies
\newline
\begin{equation}
K = C^{2} - 2\alpha A^{2}B - 2\beta AB^{2} - 2\gamma AB - \delta
B^{2} - 2\epsilon B + \frac{1}{2}\mu A^{4} + \frac{2}{3}\nu A^{3} +
\xi A^{2} + 2\zeta A.
\end{equation}
\newline
Thus K is a polynomial of order 8 in the momenta. since the
Hamiltonian H also satisfies relations (2.7) we can expect K to be a
polynomial in H and we write
\newline
\begin{equation}
K = k_{0} + k_{1}H + k_{2}H^{2} + k_{3}H^{3} + k_{4}H^{4}
\end{equation}
\newline
where $k_{0},..,k_{4}$ are constants. Note that eq. (2.9) together
with (2.8) represents a polynomial relation between the integrals H,
A, B and C in agreement with the fact that only 3 of them can be
functionally independent.
\newline
\section{Cubic Poisson algebras for classical superintegrable
systems} Let us now apply the formalism of Section 2 to all
classical superintegrable potentials that separate in Cartesian
coordinates and allow a third order integral of motion$^{22,23}$.
They have the form (1.1),...,(1.3) , so we have
$a(q_{1},q_{2})=c(q_{1},q_{2})=1$, $b(q_{1},q_{2})=0$ in (2.1).
Eight such systems exist, four of them are well-known, four first
presented in$^{23}$.
\newline
\textbf{Case 1}. The isotropic harmonic oscillator.
\newline
We have
\newline
\begin{equation}
H = \frac{P_{1}^{2}}{2} +
\frac{P_{1}^{2}}{2}+\frac{\omega^{2}}{2}(x^{2} + y^{2})
\end{equation}
\newline
The formalism of Section 2 applies, but we cannot expect it to
provide anything new. indeed, in this case we have a well known u(2)
algebra of first and second order integrals of motion:
\newline
\begin{equation}
L_{3}=xP_{2}-yP_{1},X_{1}=P_{1}^{2}+\omega^{2}x^{2},X_{2}=P_{2}^{2}+\omega^{2}y^{2}
\end{equation}
\[X_{3}=P_{1}P_{2}+\omega^{2}xy\]
\newline
we can chose
\newline
\begin{equation}
A = P_{1}^{2} - P_{2}^{2}+\omega^{2}(x^{2} - y^{2})=X_{1}-X_{2}
\end{equation}
\[ B =L_{3}X_{2}= xP_{2}^{3} -yP_{1}P_{2}-\omega^{2}y^{3}P_{1}+\omega^{2} xy^{2}P_{2} \]
\newline
we obtain
\[ \{A,B\}=C=-4X_{2}X_{3} \]
\begin{equation}
\{A,C\}= -16\omega^{2}B
\end{equation}
\[ \{B,C\}= 2A^{3} - 6HA^{2} + 8H^{3} \]
\newline
Thus, relations (2.5) are satisfied, but we remain in the enveloping
algebra of u(2). The Casimir operator of the cubic algebra in this
case is
\newline
\begin{equation}
K = C^{2} + 16\omega^{2}B^{2} + 2A^{4} - 6HA^{3} + 8H^{3}A = 16H^{4}
\end{equation}
\newline
However, in this case H is simply the central element of the u(2)
algebra (3.2) and it generates the center of the enveloping algebra
of u(2)( a hence also of the cubic algebra).
\newline
\newline
\textbf{Case 2}. A quadratically superintegrable Hamiltonian$^{5,6}$
\newline
\begin{equation}
H = \frac{P_{1}^{2}}{2} + \frac{P_{2}^{2}}{2} +
\frac{\omega^{2}}{2}(x^{2} +y^{2}) + \frac{b}{x^{2}} +
\frac{c}{y^{2}}
\end{equation}
The Hamiltonian allows two second order integrals of motion, namely
\begin{equation}
X_{1}= P_{1}^{2} - P_{2}^{2} + \omega^{2}(x^{2}-y^{2}) +
\frac{2b}{x^{2}} -\frac{2c}{y^{2}}
\end{equation}
\[X_{2}=L_{3}^{2}+2r^{2}(\frac{b}{x^{2}}+\frac{c}{y^{2}}).\]
The Hamiltonian H allows separation of variables in cartesian and
polar coordinates and also in elliptic ones. We shall identify
$X_{1}=A$ and the cubic integral B is equal to
\begin{equation}
B =-\frac{1}{8}\{X_{1},X_{2}\}=xP_{1}P_{2}^{2}-yP_{1}^{2}P_{2} + xy(
\frac{-2b}{x^{3}} + \omega^{2}x)P_{2} - xy(\frac{-2c}{y^{3}}+
\omega^{2}y)P_{1}
\end{equation}
\newline
\[ \{A,B\}=C=4(A^{2}-2HA-H^{2}+8\omega^{2}X_{2}-16\omega^{2}(b+c)\]
\begin{equation}
\{A,C\}= -64\omega^{2}B
\end{equation}
\[ \{B,C\}= -2A^{3} + 8H^{2}A + 64(c-b)\omega^{2}H + 32(c+b)\omega^{2}A \]
\begin{equation}
K = C^{2} + 64\omega^{2}B^{2} - A^{4} + (8H^{2} +
32(c+b)\omega^{2})A^{2} + 128(c-b)\omega^{2}HA
\end{equation}
\[=16H^{4} - 128\omega^{2}(c+b)H^{2} + 1024\omega^{4}bc\]
\newline
In this case there is no underlying finite-dimensional Lie algebra,
however $X_{1},X_{2}$ and H generate a quadratic Lie
algebra$^{16,18}$
\newline
\[\{X_{1},X_{2}\}=C_{q}\]
\begin{equation}
\{X_{1},C_{q}\}=-8X_{1}^{2}+16X_{1}H-16\omega^{2}X_{2}+32\omega^{2}(b+c)
\end{equation}
\[\{X_{2},C_{q}\}=16X_{1}X_{2}-16HX_{2}+32H(c-b)\]
\newline
With $C_{q}=-B$ and the relations (3.9) of the  cubic algebra are
consequences of (3.11)
\newline
\newline
\textbf{Case 3}. A further quadratically superintegrable
Hamiltonian$^{5,6}$ is
\begin{equation}
H = \frac{P_{1}^{2}}{2} + \frac{P_{2}^{2}}{2} +
\frac{\omega^{2}}{2}(4x^{2} + y^{2}) + \frac{b}{y^{2}} + cx
\end{equation}
\newline
The two second order integrals of motion are
\newline
\begin{equation}
A = P_{1}^{2} - P_{2}^{2} + \omega^{2}(4x^{2} - y^{2}) -
\frac{2b}{y^{2}} + 2cx
\end{equation}
\[X_{2}=xP_{2}^{2}-yP_{1}P_{2}+\frac{2bx}{y^{2}}-2axy^{2}-\frac{c}{2}y^{2}\]
and the Hamiltonian allows the separation of variables in cartesian
and parabolic coordinates. The third order integral is
\begin{equation}
B =-\frac{1}{4}\{X_{1},X_{2}\}=P_{1}P_{2}^{2} + (\frac{2b}{y^{2}} -
\omega^{2}y^{2})P_{1} + (4\omega^{2}xy + cy)P_{2}
\end{equation}
\newline
\[ \{A,B\}=C=16\omega^{2}X_{2} \]
\begin{equation}
\{A,C\}= -64\omega^{2}B
\end{equation}
\[ \{B,C\}= 16\omega^{2}HA + 16\omega^{2}H^{2} - 12\omega^{2}A^{2} + 8c^{2}H - 4c^{2}A
+ 128\omega^{4}b \]
\newline
\begin{equation}
 K = C^{2} + 64\omega^{2}B^{2} - 8\omega^{2}A^{3} + (16\omega^{2}H - 4c^{2})A^{2} +
(32\omega^{2}H^{2} + 16c^{2}H + 256\omega^{4}b)A
\end{equation}
\[=64\omega^{2}H^{3}+ 16c^{2}H^{2} - 512\omega^{4}bH - 128\omega^{2}bc^{2} \]
\newline
As in Case 2 we have a cubic algebra but all relations in it follow
from the already known quadratic algebra$^{16,18}$.
\newline
\newline
\textbf{Case 4}. We have
\begin{equation}
H = \frac{P_{1}^{2}}{2} + \frac{P_{2}^{2}}{2} +
\frac{\omega^{2}}{2}(9x^{2} + y^{2})
\end{equation}
Which is anisotropic harmonic oscillator with a rational frequency
ratio $\omega_{1}/\omega_{2}$=3/1. The second and third order
integrals are
\begin{equation}
A =P_{x}^{2} - P_{y}^{2} + \omega^{2}(9x^{2} - y^{2})
\end{equation}
\[ B = -yP_{1}P_{2}^{2}+xP_{2}^{3} + \frac{1}{3}\omega^{2}y^{3}P_{1} -
3\omega^{2}xy^{2}P_{2}\]
\newline
\[ \{A,B\}=C=-4P_{1}P_{2}^{3}+12\omega^{2}y^{2}P_{1}P_{2}-36\omega^{2}xyP_{2}^{2}+12\omega^{4}xy^{3}) \]
\begin{equation}
\{A,C\}= -144 \omega^{2}B
\end{equation}
\[ \{B,C\}=2A^{3} + 8H^{3} -6HA^{2} \]
\begin{equation}
K = C^{2} + 144\omega^{2}B^{2} + A^{4} - 4HA^{3} + 16AH^{3} =
16H^{4}
\end{equation}
\newline
In this case the integral B is irreducible, i.e. it is not the
Poisson bracket of lower order integrals. The cubic algebra (3.20)
is related to a u(2) invariance algebra constructed by Jauch and
Hill$^{1}$ for any anisotropic harmonic oscillator in $E_{2}$ with a
rational frequency ratio. Their Lie algebra for the ratio
$\omega_{1}/\omega_{2}$=3/1 is generated by the integrals.
\newline
\begin{equation}
\{H,\quad B_{1}=\frac{C}{H-A},\quad B_{2}=\frac{B}{H-A},\quad
B_{3}=A\}
\end{equation}
\newline
It has actually been shown that in many cases in classical mechanics
functions of integrals of motion can be constructed that generate
finite dimensional Lie algebras$^{25}$.In the case of the
anisotropic oscillator (3.17) it suffices to take the fractions
$B_{1}$ and $B_{2}$ of the polynomials (3.17),...,(3.20). The
Hamiltonians considered so far namely (3.1),(3.6),(3.12) and (3.17)
are all superintegrable, both in the classical and quantum
cases$^{23}$.
\newline
\newline
The remaining 4 cases are quite different in that the systems are
integrable only in the classical case. They are all obtained as
singular limits of quantum integrable systems. As we shall see, the
quantum systems are quite different and the potentials are expressed
in terms of Painlev\'e transcendents.
\newline
\newline
\textbf{Case 5}. The classical Hamiltonian and two integrals of
motion in this case are
\begin{equation}
H = \frac{P_{1}^{2}}{2} + \frac{P_{2}^{2}}{2} +
\beta_{1}^{2}\sqrt{|x|} + \beta_{2}^{2}\sqrt{|y|}, A =
\frac{P_{1}^{2}}{2} - \frac{P_{2}^{2}}{2} + \beta_{1}^{2}\sqrt{|x|}
- \beta_{2}^{2}\sqrt{|y|}
\end{equation}
\[ B = \beta_{2}^{4}P_{1}^{3} + \epsilon \beta_{1}^{4}P_{2}^{3} +
3\beta_{2}^{4}\beta_{1}^{2} \sqrt{|x|}P_{1} + \epsilon
3\beta_{1}^{4}\beta_{2}^{2} \sqrt{|y|}P_{2}\]
\newline
\begin{equation}
\epsilon = 1, xy > 0
\end{equation}
\[\epsilon = -1, xy < 0 \quad .\]
\newline
In the quantum case an integral of the type B exists if the
potential V(x,y)=$V_{1}(x)+V_{2}(y)$ satisfies$^{23}$
\newline
\begin{equation}
\hbar^{2}V_{1}''(x)=6V_{1}^{2}(x)-6\beta_{1}^{4}x
\end{equation}
\[\hbar^{2}V_{2}''(y)=6V_{2}^{2}(y)-6\beta_{2}^{4}y\]
\newline
The classical (and singular) limit $\hbar \rightarrow 0$ yields the
potential in (3.22). The cubic algebra (2.5) in this case simplifies
and we get
\[ \{A,B\}=C=3\beta_{1}^{4}\beta_{2}^{4} \]
\begin{equation}
\{A,C\}=0
\end{equation}
\[ \{B,C\}=0 \]
\begin{equation}
K=9\beta_{1}^{8}\beta_{2}^{8}
\end{equation}
\newline
Thus $\{A,B,C\}$ actually generate a nilpotent Lie algebra,
isomorphic to the Heisenberg algebra in one dimension (since C is
constant).
\newline
\newline
We can see from eq. (3.24) that the quantum case will be completely
different. Indeed eq. (3.24) can be solved in terms of the first
Painlev\'e transcendent $P_{I}$$^{24}$.
\newline
\newline
\textbf{Case 6}. The classical Hamiltonian and two integrals of
motion are
\begin{equation}
H = \frac{P_{1}^{2}}{2} + \frac{P_{2}^{2}}{2} +
\frac{\omega^{2}}{2}y^{2} + V(x),A = \frac{P_{1}^{2}}{2} -
\frac{P_{2}^{2}}{2} - \frac{\omega^{2}}{2}y^{2} + V(x)
\end{equation}
\[ B = -yP_{1}^{3} + xP_{1}^{2}P_{2} + (\frac{\omega^{2}}{2}x^{2} - 3V)yP_{1}-\frac{1}{\omega^{2}}(\frac{\omega^{2}}{2}x^{2}-3V)V_{x}P_{2}\]
\newline
where V satisfies a quartic equation
\newline
\begin{equation}
-9V(x)^{4} +
14\omega^{2}x^{2}V(x)^{3}+(6d-3\frac{\omega^{4}}{4}x^{4})V(x)^{2}+(\frac{3\omega^{6}}{2}x^{6}-2\omega^{2}x^{2})V(x)
\end{equation}
\[+(cx^{2}-d-d\frac{\omega^{4}}{2}x^{4}-\frac{\omega^{8}}{16}x^{8})=0\]
\newline
In the quantum case V satifies a fourth order differential
equation$^{23}$
\begin{equation}
\hbar^{2}V^{(4)}=12\omega^{2}xV'+6(V^{2})''-2\omega^{2}x^{2}V''+2\omega^{4}x^{2}
\end{equation}
that can be solved in terms of the Painlev\'e transcendent
$P_{IV}$$^{24}$. Eq. (3.27) is the solution of (3.29) for $h
\rightarrow 0$ and c and d are integration constants. In general,
eq. (3.27) has 4 roots and the expressions for them are quite
complicated. A special case occurs if $\omega^{2}$,c and d satisfy.
\newline
\begin{equation}
c=\frac{2^{3}\omega^{8}b^{3}}{3^{6}},d=\frac{\omega^{4}b^{2}}{3^{3}}
\end{equation}
\newline
where b is an arbitrary constant. Then eq. (3.28) has a double root
and we obtain
\begin{equation}
V_{1,2}(x)=\frac{\omega^{2}}{18}(2b + 5x^{2} \pm 4x\sqrt{b+x^{2}})
\end{equation}
\begin{equation}
V_{3}(x)=V_{4}(x)=(-\frac{\omega^{2}b}{3^{3}}+\frac{\omega^{2}}{2}x^{2})
\end{equation}
For V(x) satisfying (3.28) the cubic algebra is
\newline
\[ \{A,B\}=C \]
\begin{equation}
\{A,C\}=-4\omega^{2}B
\end{equation}
\[ \{B,C\}=8A^{3} + 12HA^{2} - 4H^{3} -4\frac{4b^{2}\omega^{4}}{27}A + \frac{4b^{3}\omega^{6}}{729} \]
\begin{equation}
K = C^{2}+ \omega^{2}B^{2}+4A^{4}+8HA^{3}-4bA^{2} +
(-8H^{3}+\frac{8b^{3}\omega^{6}}{729})A
\end{equation}
\[=4H^{4}-\frac{4}{27}b^{2}\omega^{4}H^{2}+\frac{8b^{3}\omega^{6}}{729}H\]
\newline
\newline
\textbf{Case 7}. The classical Hamiltonian and two integrals of
motion in this case are
\begin{equation}
H = \frac{P_{1}^{2}}{2} + \frac{P_{2}^{2}}{2} + a^{2}|y| + b^{2}
\sqrt{|x|}, A = \frac{P_{1}^{2}}{2} - \frac{P_{2}^{2}}{2} - a^{2}|y|
+ b^{2}\sqrt{|x|}
\end{equation}
\[ B = P_{1}^{3} + 3b^{2}\sqrt{|x|}P_{1}+\epsilon \frac{3b^{4}}{2a^{2}}P_{2}\]
\newline
\begin{equation}
\epsilon = -1, xy > 0
\end{equation}
\[\epsilon = 1, xy < 0 \quad .\]
\newline
In the quantum case the potential is V(x,y)=ay + V(x) where V(x)
satisfies
\begin{equation}
\hbar^{2}V''=6V^{2}-6b^{4}x ,\qquad b \neq 0
\end{equation}
This is solved in terms of the first Painlev\'e transcendent for
$\hbar \neq 0$. For $\hbar =0$ the singular limit is V=$b\sqrt{x}$.
The cubic algebra in this case is very simple and reduces to a Lie
algebra, the Heisenberg algebra as in Case 5. We have
\[  \{A,B\}=C=3b^{4} \]
\begin{equation}
\{A,C\}=0
\end{equation}
\[  \{B,C\}=0  \]
\begin{equation}
K=9b^{8}
\end{equation}
\newline
\textbf{Case 8}. In this case we have
\begin{equation}
H = \frac{P_{1}^{2}}{2} + \frac{P_{2}^{2}}{2} + ay + V(x),\quad A =
\frac{P_{1}^{2}}{2} - \frac{P_{2}^{2}}{2} - ay + V(x)
\end{equation}
\[ B = aP_{1}^{3}-bP_{1}^{2}P_{2} + a(3V(x)-bx)P_{1}-2bV(x)P_{2}\]
\newline
where V(x) satisfies a cubic equation
\begin{equation}
V(x)^{3} - 2bxV(x)^{2} + b^{2}x^{4}V(x)-d=0
\end{equation}
This is again obtained from a singular limit of a second order
nonlinear ODE in the quantum case. It is, for $\hbar \neq 0$, solved
in terms of the Painlev\'e transcendent $P_{II}$. The cubic algebra
in this case reduces to
\newline
\[  \{A,B\}=C=2ab(A+H) \]
\begin{equation}
\{A,C\}=0
\end{equation}
\[  \{B,C\}=-4a^{2}b^{2}(A+H) \]
\begin{equation}
K= 4a^{2}b^{2}H^{2}
\end{equation}
\newline
We see that (3.42) is actually a Lie algebra with centre H. If we
put
\begin{equation}
A_{1}=\frac{H+A}{2},B_{1}=-\frac{1}{2ab}B
\end{equation}
We obtain the solvable decomposable Lie algebra
\begin{equation}
\{H_{1},B_{1}\}=B_{1},\{H_{1},H\}=\{B_{1},H\}=0
\end{equation}
The cubic equation (3.41) can be taken to standard form by putting
\begin{equation}
V(x)=y(x)+\frac{2bx}{3}
\end{equation}
We obtain
\begin{equation}
y^{3}+3py+2q=0,\quad p=-(\frac{bx}{3})^{2},\quad
q=(\frac{bx}{3})^{3}-\frac{d}{2}
\end{equation}
This discriminant is
\begin{equation}
D=q^{2}+p^{3}=\frac{d^{2}}{4}-\frac{b^{3}dx^{3}}{27}
\end{equation}
\newline
For D < 0,i.e. $x^{3} > \frac{27d}{4b^{2}}$ with d<0 or $x^{3} <
\frac{27d}{4b^{2}}$ with d>0 it has 3 real roots,
\newline
For D > 0,i.e. $x^{3} < \frac{27d}{4b^{2}}$ with d<0 or $x^{3} >
\frac{27d}{4b^{2}}$ with d>0 it has 1 real root $y_{1}$ and 2
complex ones,$y_{3}=\bar y_{2}$.
\newline
\newline
The corresponding potentials are
\newline
\begin{equation}
V_{1}(x)=\frac{2bx}{3}+\frac{2^{1/3}b^{2}x^{2}}{3(2+d-2b^{3}x^{3}+3\sqrt{3}\sqrt{27d^{2}-4b^{3}dx^{3}}
)^{1/3}}
\end{equation}
\[+ \frac{(27d-2b^{3}x^{3}+3\sqrt{3}\sqrt{27d^{2}-4b^{3}dx^{3}})^{1/3}}{3 2^{1/3}}       \]
\begin{equation}
V_{2}(x)=\frac{2bx}{3}+\frac{(1+i\sqrt{3})b^{2}x^{2}}{3
2^{2/3}(2+d-2b^{3}x^{3}+3\sqrt{3}\sqrt{27d^{2}-4b^{3}dx^{3}}
)^{1/3}}
\end{equation}
\[+ \frac{(1-i\sqrt{3})(27d-2b^{3}x^{3}+3\sqrt{3}\sqrt{27d^{2}-4b^{3}dx^{3}})^{1/3}}{6 2^{1/3}}
\]
\begin{equation}
V_{3}(x)=\frac{2bx}{3}+\frac{(1-i\sqrt{3})b^{2}x^{2}}{3
2^{2/3}(2+d-2b^{3}x^{3}+3\sqrt{3}\sqrt{27d^{2}-4b^{3}dx^{3}}
)^{1/3}}
\end{equation}
\[+ \frac{(1+i\sqrt{3})(27d-2b^{3}x^{3}+3\sqrt{3}\sqrt{27d^{2}-4b^{3}dx^{3}})^{1/3}}{6 2^{1/3}}       \]
\newline
Multiple roots occur for D=0.This happens (for x $\neq$ const) for
d=0 only and then we have
\begin{equation}
V=bx
\end{equation}
as a double root(and V=0 as the simple one). We can consider the
particular Hamiltonian,
\begin{equation}
H = \frac{P_{1}^{2}}{2} + \frac{P_{2}^{2}}{2} + a^{2}|y| + b^{2}|x|,
A = \frac{P_{1}^{2}}{2} - \frac{P_{2}^{2}}{2} - a^{2}|y| + b^{2}|x|
\end{equation}
\[ B = a^{2}P_{1}^{3}-b^{2}P_{1}^{2}P_{2} + 2a^{2}b^{2}|x|P_{1}-2b^{4}|x|P_{2}\]
\newline
\section{Trajectories for classical superintegrable systems}
If a Hamiltonian system is maximally superintegrable and satisfies
certain analyticity properties, then all of its bounded trajectories
are closed and the motion is periodic$^{26}$. Here we shall discuss
the trajectories for all the 8 systems of Section 3. The
trajectories can be obtained in a uniform manner directly from the
integrals of motion. Indeed in all cases we can put
\begin{equation}
\frac{1}{2}P_{1}^{2}+f(x)=E_{1}
\end{equation}
\begin{equation}
\frac{1}{2}P_{2}^{2}+g(y)=E_{2}
\end{equation}
\begin{equation}
B=\mu P_{1}^{3}+ \nu P_{1}^{2}P_{2} + \rho P_{1}P_{2}^{2}+ \sigma
P_{2}^{3} + \phi P_{1} + \psi P_{2} = k \quad .
\end{equation}
In (4.3) $\mu$ ,$\nu$ ,$\rho$ and $\sigma$ are low order polynomials
in x and y, $\phi$ and $\psi$ are functions of x and y (all of them
known). The constants $E_{1}$ and $E_{2}$ are positive, k arbitrary.
From (4.1) and (4.2) we obtain $P_{1}^{2}$ and $P_{2}^{2}$ in term
of x and y, respectively. From eq(4.3) we obtain
\begin{equation}
( P_{1}^{2}(\mu P_{1}^{2} + \rho P_{2}^{2}+\phi)^{2} - P_{2}^{2}(\nu
P_{1}^{2}+\sigma P_{2}^{2}+\psi)^{2})^{2}+k^{4}
\end{equation}
\[= 2k^{2}(P_{1}^{2}(\mu P_{1}^{2} + \rho P_{2}^{2}+\phi)^{2}+P_{2}^{2}(\nu
P_{1}^{2}+\sigma P_{2}^{2}+\psi)^{2}) \quad .\]
\newline
Substituting for $P_{1}^{2}$ and $P_{2}^{2}$ from (4.1) and (4.2) we
obtain the equation for the trajectories. Directly from eq (4.1) and
(4.2) we see that the motion is bounded if there exist two
constants, $x_{o}$ and $y_{o}$,such that we have
\begin{equation}
f(x) \geq 0 ,\quad x^{2} > x_{o}^{2} ,\quad g(y) \geq 0,
y^{2}>y_{o}\quad .
\end{equation}
This direct method of computing trajectories is universal, hence not
necessarily convenient in any special case. For instance for the
harmonic oscillator it gives a polynomial of order 16 in x and y
which can then be simplified to a second order one. However it
avoids the problem of integrating the equation of motion and then
eliminating time.
\newline
\newline
\textbf{Case 1}. The trajectories of the harmonic oscillator are
well known and are ellipses.
\newline
\newline
\textbf{Case 2}. The trajectories are given in$^{6}$ and are easily
obtained by integrating the equation of motion:
\newline
\begin{equation}
x(t)^{2}= \frac{E_{1}}{\omega^{2}} +
\sqrt{\frac{E_{1}^{2}}{2\omega^{4}} -
\frac{2b}{\omega^{2}}}sin(2\omega t+c_{1})
\end{equation}
\begin{equation}
 y(t)^{2}= \frac{E_{2}}{\omega^{2}} + \sqrt{\frac{E_{2}^{2}}{2\omega^{4}} -
\frac{2c}{\omega^{2}}}sin(2\omega t+c_{2})
\end{equation}
\newline
\textbf{Case 3}. The trajectories are also given in$^{6}$ and are
obtained as in Case 2.
\newline
\begin{equation}
 x(t)^{2}= \frac{-c}{4\omega^{2}} + \sqrt{\frac{E_{1}}{\omega^{2}} +
\frac{c^{2}}{32\omega^{2}}}sin(2 \omega t+c_{1})
\end{equation}
\begin{equation}
 y(t)^{2}= \frac{E_{2}}{\omega^{2}} + \sqrt{\frac{E_{2}^{2}}{\omega^{4}} -
\frac{2c}{\omega^{2}}}sin(2 \omega t+c_{2})
\end{equation}
\newline
\textbf{Case 4}.The Case of the anisotropic oscillator is well
known. The trajectories are what is called Lissajous figures.
\newline
\newline
\begin{equation}
 x(t)=\sqrt{\frac{2E_{1}}{9 \omega^{2}}}sin(3 \omega t+c_{1})
 \end{equation}
\begin{equation}
y(t)=\sqrt{\frac{2E_{2}}{\omega^{2}}}sin(\omega t+c_{2})
\end{equation}
\newline
\newline
\textbf{Case 5}. The potential is
\begin{equation}
V = \beta_{1}^{2}\sqrt{|x|}+\beta_{2}^{2}\sqrt{|y|} \quad .
\end{equation}
\newline
We find the trajectories using the equation (4.4). The bounded
trajectories are closed. Examples are shown on Fig 5.
\newline
\newline
\textbf{Case 6}. We know from the Section 3 that the only particular
case where two roots coincide is (3.30). The trajectories for the
potentials $V_{1,2}(x)$ satisfy
\newline
\begin{equation}
y(t)=\sqrt{\frac{2E_{2}}{\omega^{2}}}sin(\omega t+c_{2})
\end{equation}
\begin{equation}
\int \frac{dx}{\sqrt{-\frac{10\omega^{2}}{9}x^{2} \mp
\frac{4\omega^{2}}{9}x\sqrt{b + x^{2} }+ 2E_{1} }} = t + c_{1}
\end{equation}
The trajectories were obtained numerically directly from the
equations of motion. We have closed trajectories and examples are
shown on Fig 6. For $V_{3}(x)$ the potential is a shifted harmonic
oscillator, so the trajectories are ellipses.
\newline
\newline
\textbf{Case 7}. The potential is
\begin{equation}
V = a^{2}|y|+ b^{2}\sqrt{|x|}
\end{equation}
\newline
We find the trajectories using equation (4.4). The bounded
trajectories are closed. They are shown on Fig 7.
\newline
\newline
\textbf{Case 8}. There is the particular case
\begin{equation}
V = a^{2}|y|+ b^{2}|x|
\end{equation}
We use eq. (4.4) to calculate the trajectories. Again the bounded
trajectories are closed. An example is shown on Fig 8.
\newpage
\section{Conclusion}
The main results of this article are the study of the algebras of
integrals of motion, presented in Sections 2 and 3 and the
investigation of classical trajectories in Section 4.
\newline
In all 8 cases of superintegrable systems, separating in cartesian
coordinates and allowing a third order integral of motion, the
integrals of motion generate a cubic Poisson algebra. In many cases
this polynomial algebra is reducible that is it is a consequence of
the existence of a simpler algebraic structure.
\newline
In 4 cases the simplest underlying structure is a Lie algebra. Thus
in Case 1, the isotropic harmonic oscillator, the integrals of
motion generate the u(2) algebra with the Hamiltonian as its central
element. The third order operator B and the fourth order one C lie
in the enveloping algebra of u(2).
\newline
In Case 5 the potential
$V(x,y)=\beta_{1}^{2}\sqrt{|x|}+\beta_{2}^{2}\sqrt{|y|}$ is
nonanalytical at the origin. In this case the Poisson bracket of the
second and the third order integrals is a constant. Hence A,B and C
generate the Heisenberg Lie algebra.
\newline
Case 7 with $V(x,y)=a^{2}|y| + b^{2}\sqrt{|x|}$ is similar to Case 5
in that A, B and C generate the Heisenberg Lie algebra (with C
constant). In both of these cases the Lie algebra is actually
\newline
\begin{equation}
\{H\}\bigoplus \{A,B,C\}
\end{equation}
i.e. the direct sum of a one-dimensional Lie algebra, generated by
the Hamiltonian, and the Heisenberg algebra.
\newline
In Case 8 we have V(x,y)= ay + V(x) with V(x) satisfying the cubic
equation (3.41), the integrals generate a solvable decomposable Lie
algebra (3.45). The Hamiltonian H commutes with the elements of the
two-dimensional solvable algebra.
\newline
The Case 2 and 3 are different. The Hamiltonians (3.6) and (3.12)
are actually quadratically superintegrable and the third order
integral B is the Poisson bracket of two second order ones. These
second order integrals give rise to a quadratic Poisson
algebra$^{15,16}$ and the cubic algebra of our Section 3 is an
algebraic consequence of the quadratic one.
\newline
Finally, two irreducible cases remain. One is the Case 4, the
anisotropic harmonic oscillator (3.17). We have a genuine third
order polynomial algebra (3.19), (3.20) of polynomial integrals
H,A,B and C of order 2,2,3 and 4, respectively. Somewhat
artificially, we can construct the Lie algebra u(2) of eq. (3.21),
using rational functions of these integrals.
\newline
Case 6 is again irreducible, i.e. we obtain the genuinely cubic
polynomial algebra (3.33), (3.34), again involving two second order,
one third order and one fourth order integral. The condition for
this algebra to close is precisely that V(x) should satisfy the
quartic equation (3.28).
\newline
To our knowledge, no classification of polynomial algebras exist,
not even of quadratic ones, still less of cubic algebras. We can
however see that the linear parts of the algebras of Case 4 and Case
6 are not isomorphic as Lie algebras. That of Case 4 is solvable,
that of Case 6 is simple.
\newline
Trajectories are studied in Section 4 with examples on Fig 1,...,8.
The most important result is that all finite trajectories are
closed. We note that the potentials of Case 1 and 4 are analytic
everywhere, so superintegrability implies periodic motion$^{26}$. In
cases 2 and 3 the potentials are singular along one or both
coordinate axes, analytic elsewhere. In Case 5 the potential is
nonanalytic with nonisolated branchpoints along the axes. That not
withstanding, bounded trajectories turn out to be closed.
\newline
The functions $V_{1,2}(x)$ in Case 6 are nonanalytic for $x^{2}=-b$
for b>0. Similar statements apply for Case 7. In Case 8 we have
studied a psecial case when the cubic equation (3.41) has a double
root. In this case we either have a linear potential $V(x,y)=\alpha
x + \beta y$ with no bounded trajectories, or the nonanalytical one
$V(x,y)=a^{2}|y|+b^{2}|x|$ (4.16) that we have investigated.
\newline
Our conclusion is that the bounded trajectories are always closed
for the superintegrable potentials, wheter they are analytical, or
not.
\newline
An investigation of the properties of superintegrable systems with a
third order integral of motion is in progress.
\newline
\newline
\textbf{Acknowledgments} The authors thanks Frederick Tremblay for
useful discussions. I.M. benefited from an FQRNT student fellowship.
\newline
\newline
\section{\textbf{References}}
$^{1}$J.M Jauch and E.L Hill,On the problem of degeneracy in quantum
mechanics,Phys Rev,57,641-645 (1940).
\newline
\newline
$^{2}$M.Moshinsky and Yu F.Smirnov,The Harmonic Oscillator In Modern
Physics,Harwood,Amsterdam,641-645 (1966).
\newline
\newline
$^{3}$V. Fock, Zur Theorie des Wasserstoffsatoms Phys.98,145-154
(1935).
\newline
\newline
$^{4}$V. Bargmann,Z. Zur Theorie des Wasserstoffsatoms Phys.
99,576-582 (1936).
\newline
\newline
$^{5}$J. Fris, V. Mandrosov, Ya.A. Smorodinsky, M. Uhlir and P.
Winternitz, On higher symmetries in quantum mechanics, Phys. Lett.
16, 354-356 (1965).
\newline
\newline
$^{6}$P.Winternitz, Ya.A Smorodinsky, M.Uhlir and I.Fris, Symmetry
group in classical and quantum mechanics ,
J.Nucl.Phys.(U.SS.R)4,625-635 (1966).
\newline
\newline
$^{7}$A. Makarov, Ya.A. Smorodinsky, Kh. Valiev and P. Winternitz,A
systematic search for nonrelativistic systems with dynamical
symmetries, Nuovo Cim. A52, 1061-1084 (1967).
\newline
\newline
$^{8}$E.G. Kalnins,J.M.Kress and W.Miller Jr, Second-order
superintegrable systems in conformally flat spaces. I,II,III,
J.Math.Phys,46,053509 28 pages,053510 15 pages,103507 28
pages,(2005).
\newline
\newline
$^{9}$E.G. Kalnins, W. Miller Jr. and G. S. Pogosyan. Completeness
of multiseparability in E2,C. J. Phys. A33,4105-4120 (2000).
\newline
\newline
$^{10}$M.B.Sheftel, P.Tempesta and P.Winternitz,Exact solvability of
superintegrable systems, J.Math.Phys.42,659-673 (2001).
\newline
\newline
$^{11}$P.Tempesta, A.V Turbiner and P.Winternitz,Exact solvability
of superintegrable systems, J.Math.Phys.42,4248-4257 (2001).
\newline
\newline
$^{12}$N.W. Evans. Superintegrability in classical mechanics. Phys.
Rev. A 41,5666-5676 (1990).
\newline
\newline
$^{13}$N.W. Evans. Group theory of the Smorodinsky-Winternitz
system. J. Math. Phys. 32,3369-3375 (1991).
\newline
\newline
$^{14}$Ya. I Granovskii,A.S. Zhedanov  and I.M. Lutzenko , Quadratic
Algebra as a Hidden Symmetry of the Hartmann Potential, J. Phys. A24
3887-3894 (1991).
\newline
\newline
$^{15}$P.Létourneau and L.Vinet, Superintegrable systems:Polynomial
algebras and quasi-exactly solvable hamiltonians,Ann. Physics
243,144-168(1995).
\newline
\newline
$^{16}$C.Daskaloyannis,Quadratic poisson algebras of two-dimensional
classical superintegrable systems and quadratic associative algebras
of quantum superintegrable systems, J.Math.Phys.42,1100-1119,
(2001).
\newline
\newline
$^{17}$D.Bonatsos,C.Daskaloyannis and K.Kokkotas, Deformed
oscillator algebras for two-dimensional quantum superintegrable
systems.  Phys. Rev.A 50,3700-3709 (1994).
\newline
\newline
$^{18}$D.Bonatsos,C.Daskaloyannis and K.Kokkotas, Quantum-algebraic
description of quantum superintegrable systems in two dimensions.
Phys. Rev. A 48,R3407-R3410. 81S05 (1993).
\newline
\newline
$^{19}$J. Drach Sur l'intégration logique des équations de la
dynamique à deux variables:Forces conservatrices.Intégrales
cubiques.Mouvements dans le plan,C.R. acad. Sci III,200,22-26
(1935).
\newline
\newline
$^{20}$J. Drach, Sur l'intégration logique et sur la transformation
des équations de la dynamique à deux variables:Forces
conservatrices.Intégrales.C.R.Acad.Sci III,599-602 (1935).
\newline
\newline
$^{21}$J.Hietarinta,Classical vs. Quantum
integrability,J.Math.Phys.25,1833-1840 (1984).
\newline
\newline
$^{22}$S.Gravel and P.Winternitz, Superintegrability with
third-order integrals in quantum and classical mechanics J. Math.
Phys. 43 5902-5912 (2002).
\newline
\newline
$^{23}$S.Gravel, Hamiltonians separable in Cartesian coordinates and
thirdorder integrals of motion, J. Math. Phys. 45 1003-1019 (2004).
\newline
\newline
$^{24}$E.L.Ince,Ordinary Differential Equations,Dover,New York,
(1956).
\newline
\newline
$^{25}$G. Gonora,P. Kosinski,M. Majenski and P.Mashenka,Symmetry
algebras for superintegrable systems.J.Phys.A,39,343-349 (2006).
\newline
\newline
$^{26}$N.N.Nekhoroshev,Action-angle variables and their
generalizations.Trudy Mskov.Mat.Obshch.26,181-198 (1972).
\newpage
\textbf{Figure captions}
\newline
\newline
Fig1.A trajectory for $V=\frac{\omega^{2}}{2}(x^{2}+y^{2})$.
Parameter $\omega^{2}=2$, $v_{xo}=-1.5$, $x_{o}=5$, $v_{yo}=-1.2$,
$y_{o}=-2$, t=[0,400]
\newline
\newline
Fig2.A trajectory for $\frac{\omega^{2}}{2}(x^{2} +y^{2}) +
\frac{b}{x^{2}} + \frac{c}{y^{2}}$. Parameter $\omega^{2}=2$,b=2 and
c=3, $v_{xo}=-1.5$, $x_{o}=5$, $v_{yo}=-1.2$, $y_{o}=-2$, t=[0,400]
\newline
\newline
Fig3.A trajectory for $\frac{\omega^{2}}{2}(4x^{2} + y^{2}) +
\frac{b}{y^{2}} + cx$. Parameter $\omega^{2}=1$,b=2 and c=3,
$v_{xo}=-1.5$, $x_{o}=5$, $v_{yo}=-1.2$, $y_{o}=-2$, t=[0,400]
\newline
\newline
Fig4.A trajectory for $\frac{\omega^{2}}{2}(9x^{2} + y^{2})$.
Parameter $\omega^{2}=1$ and d=3, $v_{xo}=-1.5$, $x_{o}=5$,
$v_{yo}=-1.2$, $y_{o}=-2$, t=[0,400]
\newline
\newline
Fig5.A trajectory for $V =
\beta_{1}^{2}\sqrt{|x|}+\beta_{2}^{2}\sqrt{|y|}$. Parameter
$E_{1}=E_{2}=k=1$
\newline
\newline
Fig6.A trajectory for $\frac{\omega^{2}}{18}(2b + 5x^{2} \pm
4x\sqrt{b+x^{2}})$.  Parameter $\omega^{2}=1$ and d=3,
$v_{xo}=-1.5$, $x_{o}=5$, $v_{yo}=-1.2$, $y_{o}=-2$, t=[0,400]
\newline
\newline
Fig7.A trajectory for $V = a^{2}|y|+ b^{2}\sqrt{|x|}$. Parameter
$E_{1}=E_{2}=k=1$
\newline
\newline
Fig8.A trajectory for $V = a^{2}|y|+ b^{2}|x|$. Parameter
$E_{1}=E_{2}=k=1$
\newpage
\centerline{ \epsfxsize=4in \epsfysize=4in \epsfbox{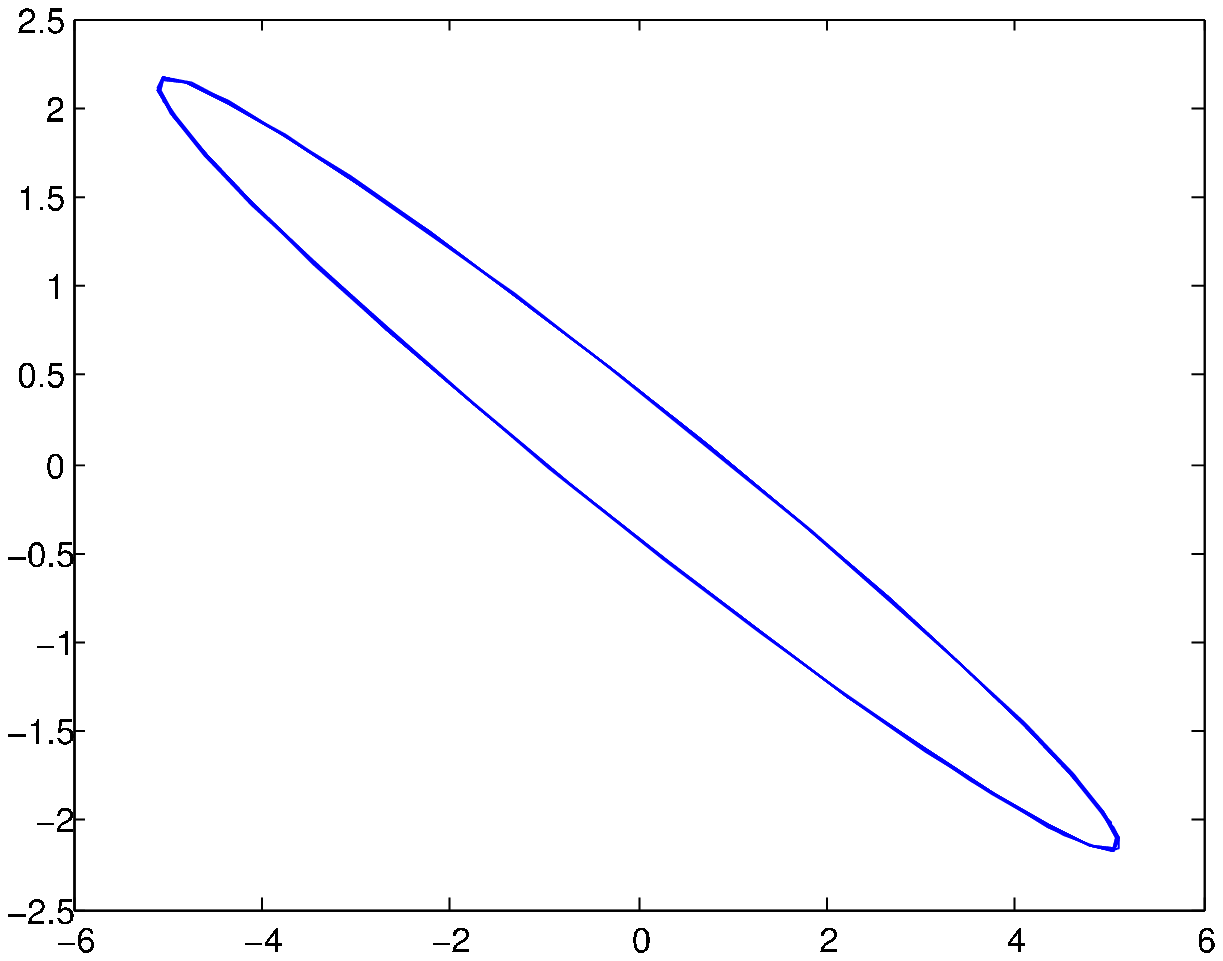} }
\newpage
\centerline{ \epsfxsize=4in \epsfysize=4in \epsfbox{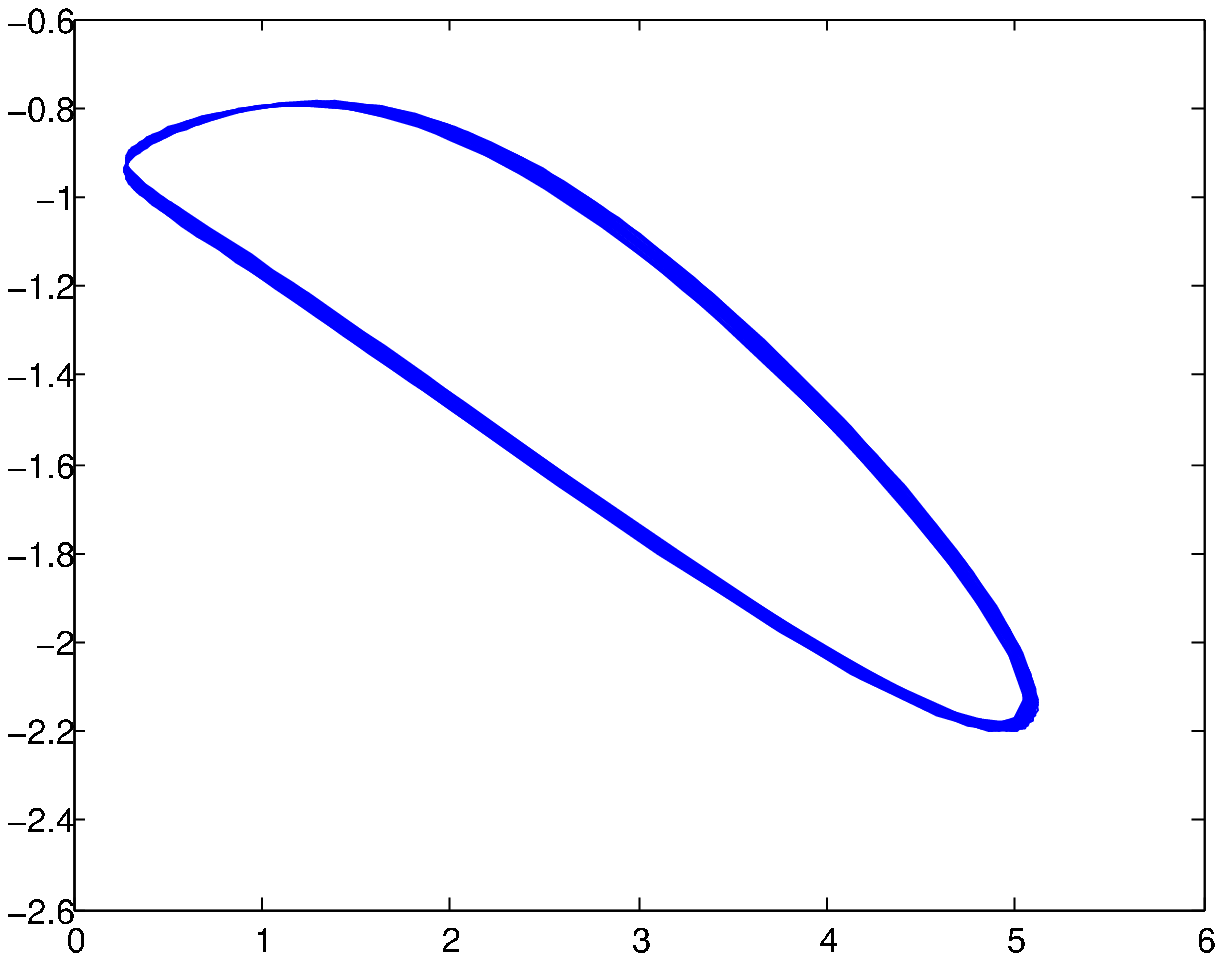} }
\newpage
\centerline{ \epsfxsize=4in \epsfysize=4in \epsfbox{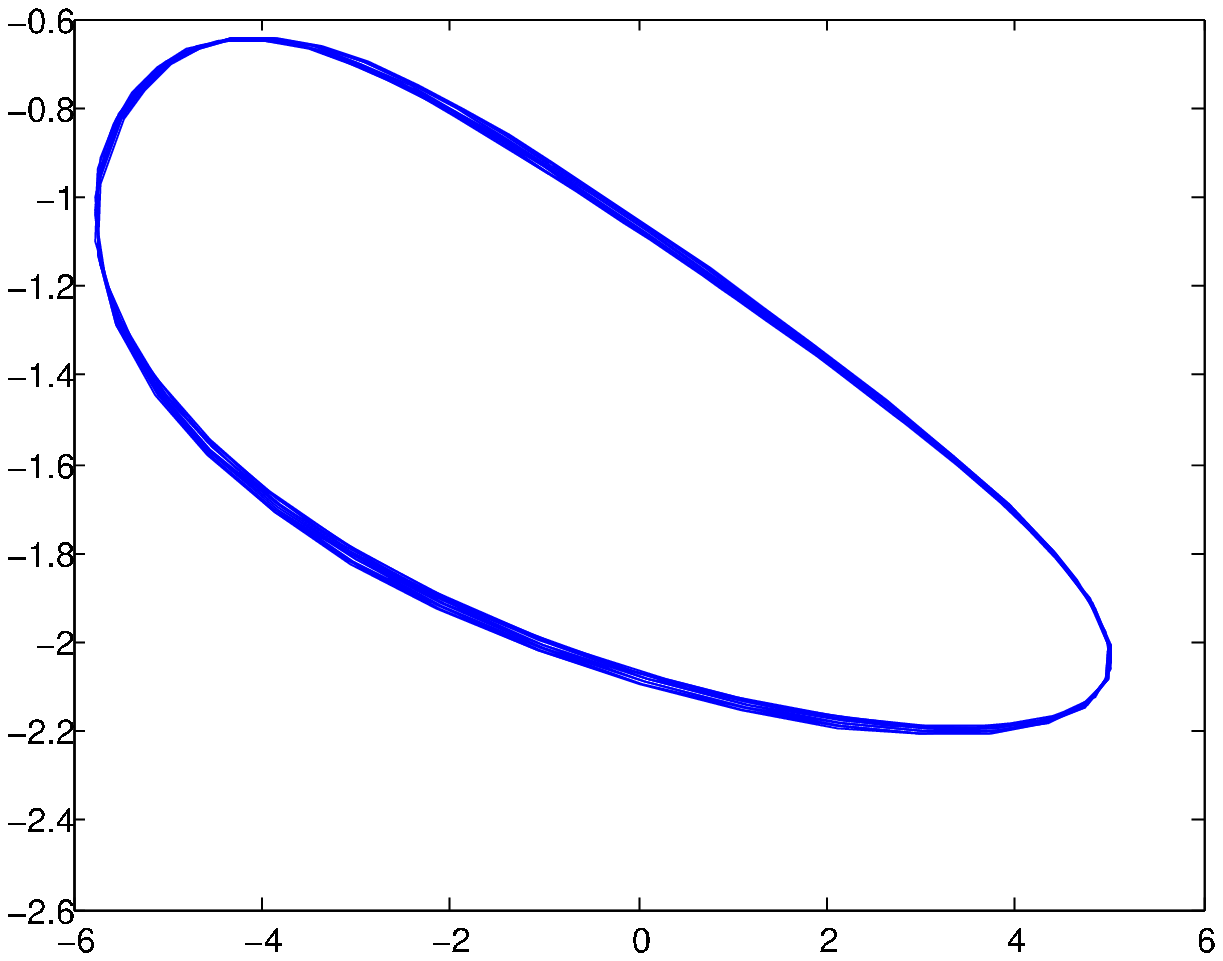} }
\newpage
\centerline{ \epsfxsize=4in \epsfysize=4in \epsfbox{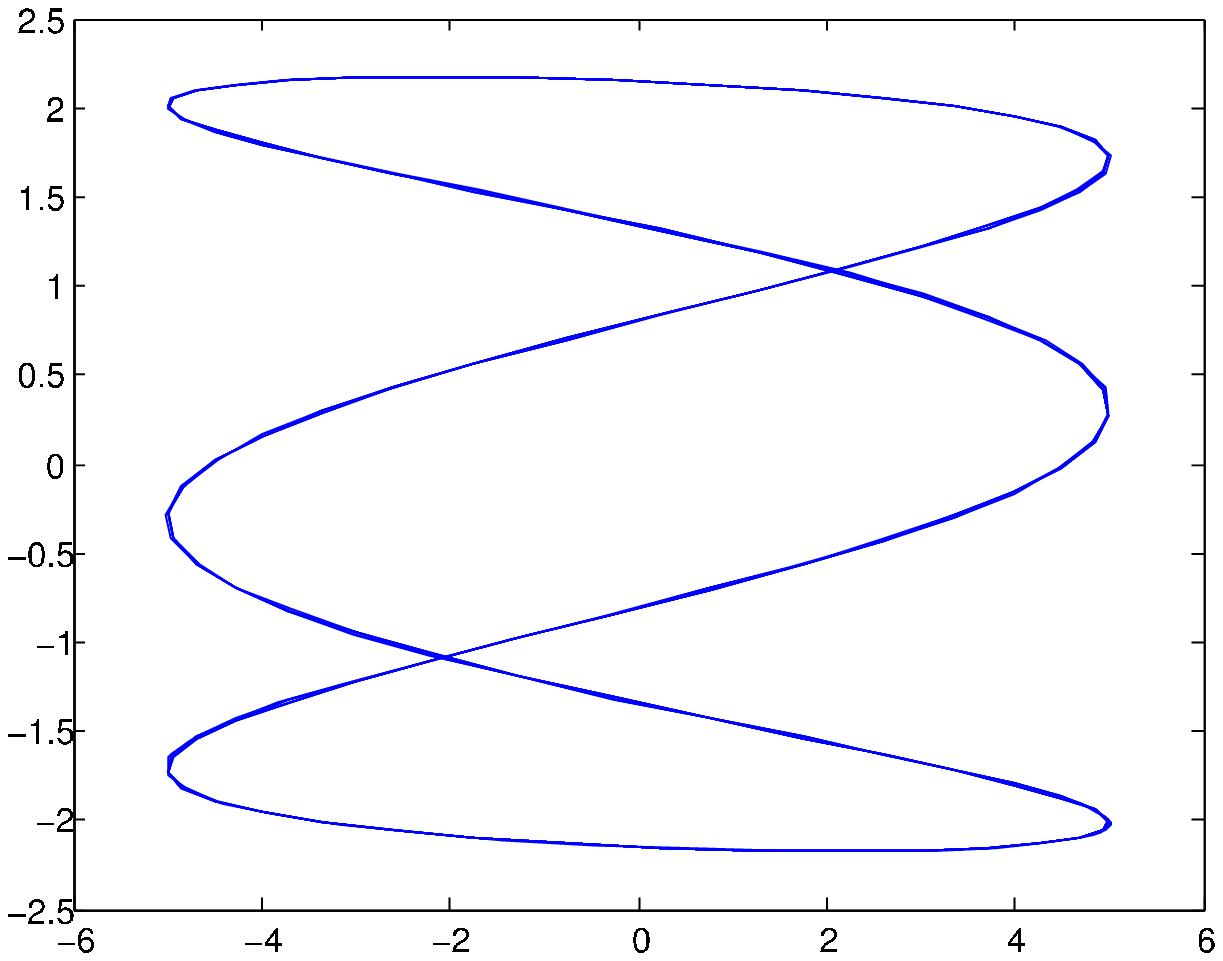} }
\newpage
\centerline{ \epsfxsize=4in \epsfysize=4in \epsfbox{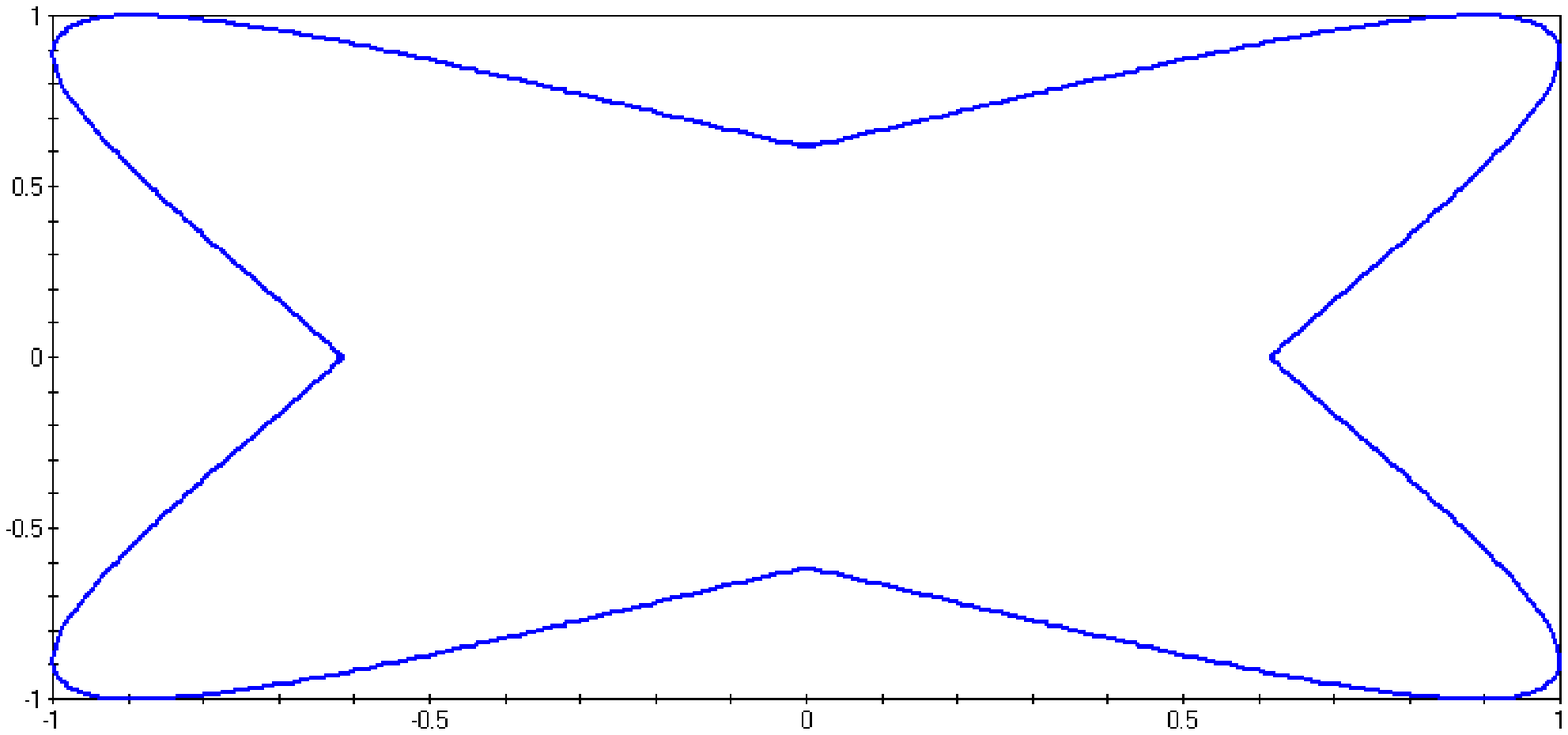} }
\newpage
\centerline{ \epsfxsize=4in \epsfysize=4in \epsfbox{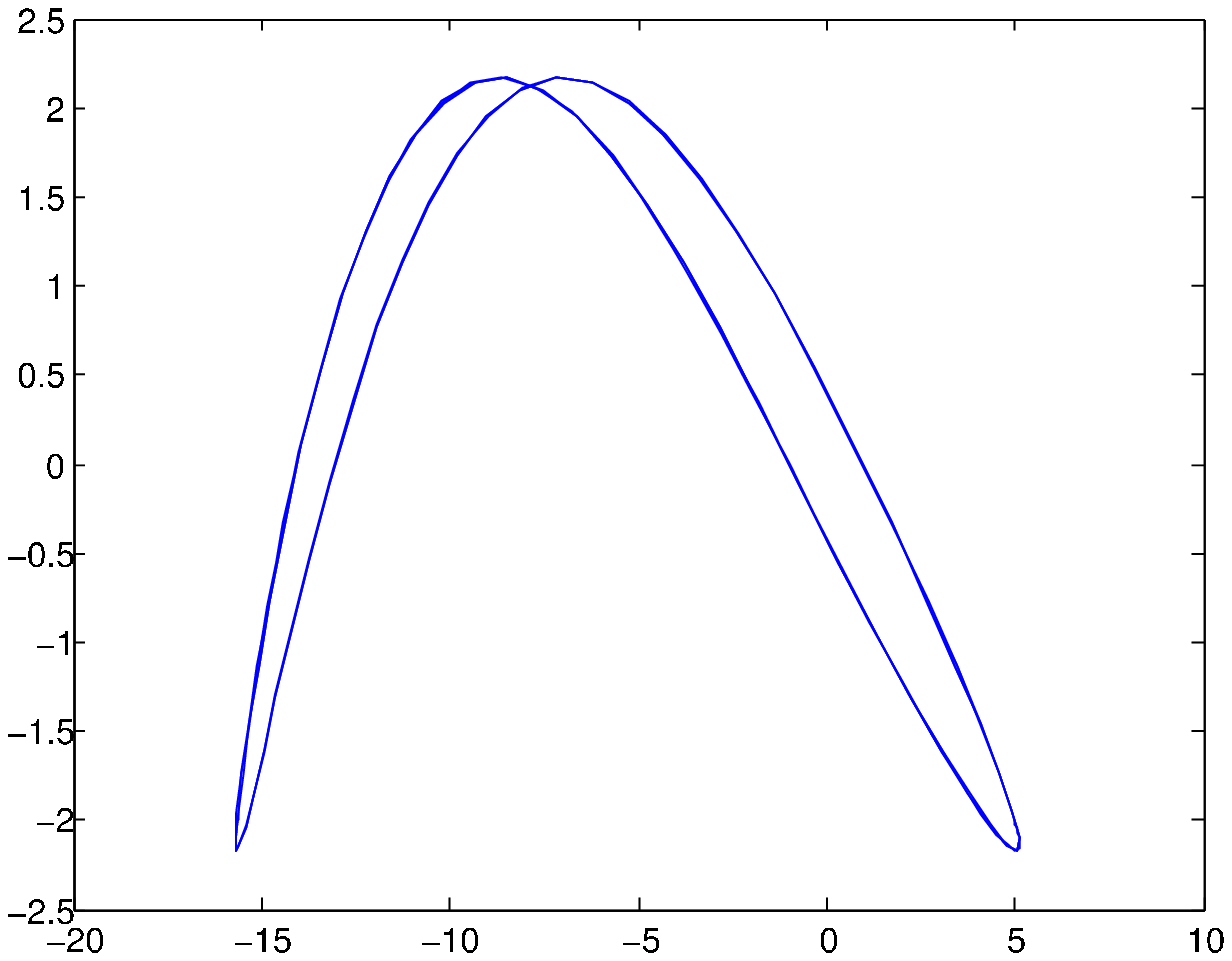} }
\newpage
\centerline{ \epsfxsize=4in \epsfysize=4in \epsfbox{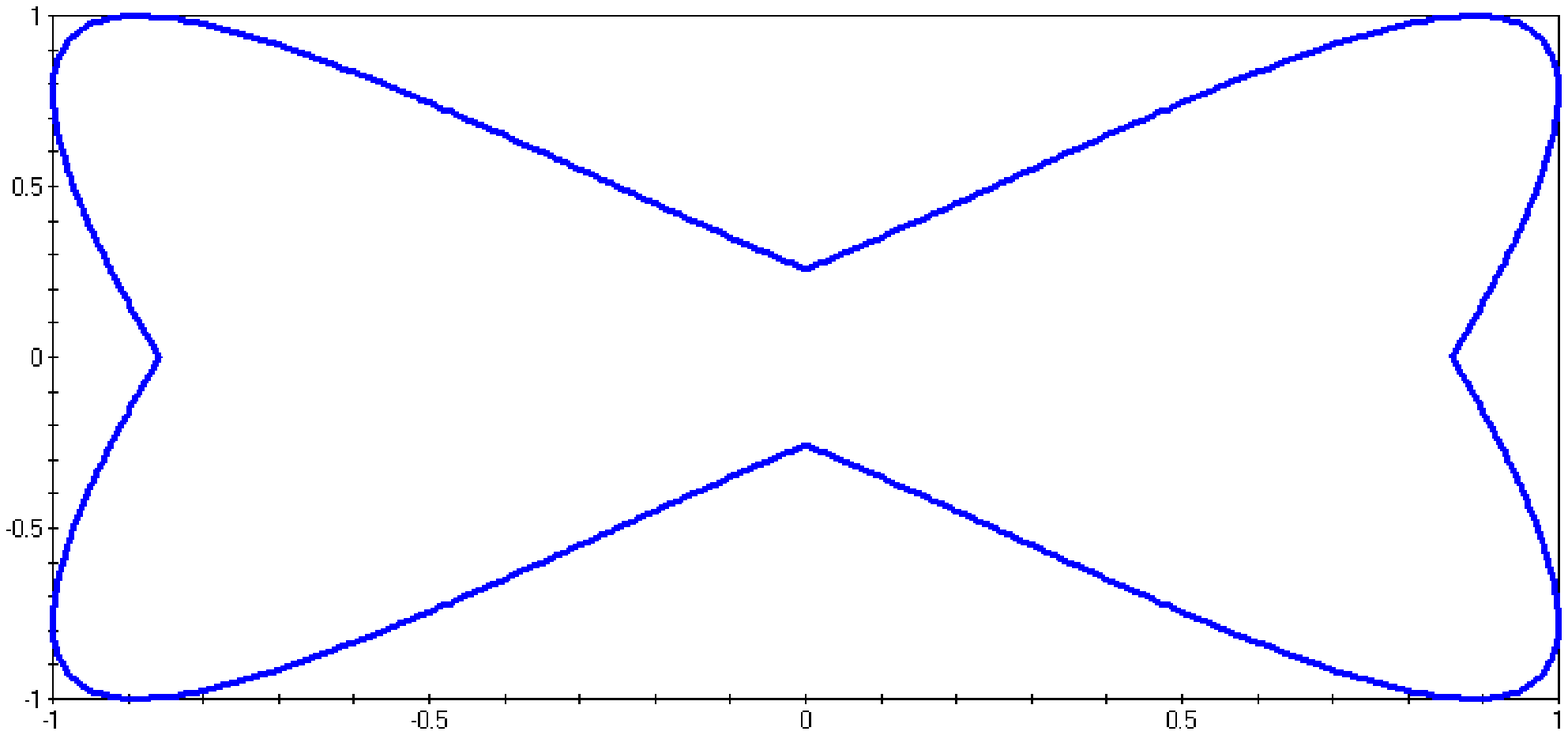} }
\newpage
\centerline{ \epsfxsize=4in \epsfysize=4in \epsfbox{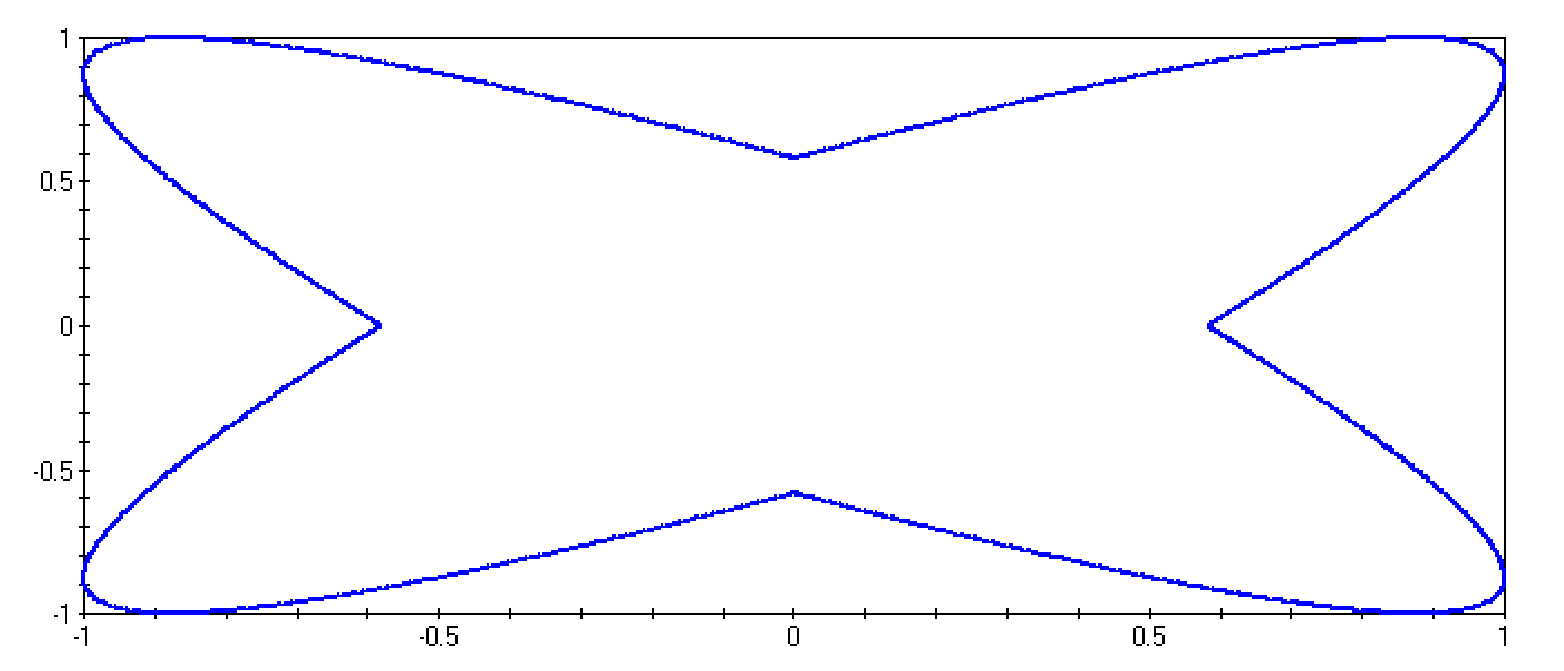} }

\end{flushleft}
\end{document}